\documentclass[12pt]{article}
\usepackage{cite}
\newcommand{\state}[1]{\left| \right. \left. #1 \right>}
\newcommand{\ba}{\begin{array}}
\newcommand{\ea}{\end{array}}
\newcommand{\be}{\begin{equation}}
\newcommand{\ee}{\end{equation}}
\newcommand{\bea}{\begin{eqnarray}}
\newcommand{\eea}{\end{eqnarray}}

\newcommand{\la}{\langle}
\newcommand{\ra}{\rangle}

\newcommand{\p}{\partial}

\newcommand{\bin}[2]{\left( \begin{array}{c} #1 \\ #2 \end{array} \right)}
\newcommand{\qbin}[2]{\left[ \begin{array}{c} #1 \\ #2 \end{array} \right]}

\def\CP{{\mathcal{P}}}
\def\CQ{{\mathcal{Q}}}

\def\IB{\relax\hbox{$\inbar\kern-.3em{\rm B}$}}
\def\IC{\relax\hbox{$\inbar\kern-.3em{\rm C}$}}
\def\ID{\relax\hbox{$\inbar\kern-.3em{\rm D}$}}
\def\IE{\relax\hbox{$\inbar\kern-.3em{\rm E}$}}
\def\IF{\relax\hbox{$\inbar\kern-.3em{\rm F}$}}
\def\IG{\relax\hbox{$\inbar\kern-.3em{\rm G}$}}
\def\IGa{\relax\hbox{${\rm I}\kern-.18em\Gamma$}}
\def\IH{\relax{\rm I\kern-.18em H}}
\def\IK{\relax{\rm I\kern-.18em K}}
\def\IL{\relax{\rm I\kern-.18em L}}
\def\IP{\relax{\rm I\kern-.18em P}}
\def\IR{\relax{\rm I\kern-.18em R}}
\def\KZ{Knizhnik-Zamolodchikov }
\def\IZ{\relax{\rm Z\kern-.5em Z}}

\def\half{\frac{1}{2}}
\def\p{\partial}
\def\f{\frac}

\def\ba{\begin{array}}
\def\ea{\end{array}}

\begin{document}

\begin{titlepage}


\begin{flushright}

\end{flushright}

\vskip 2 cm

\begin{center}
{\LARGE The origin of multiplets of chiral fields in $SU(2)_k$ at rational level}
\vskip 1 cm

{\large A. Nichols\footnote{Present address:\\ {\it Physikalisches Institut der Universit\"at Bonn,\\}{\it Nu\ss allee 12, D-53115 Bonn, Germany.}}\footnote{nichols@th.physik.uni-bonn.de}}

\begin{center}
{\em National Research Centre Demokritos, \\
Institute of Nuclear Physics, \\
Agia Paraskevi, \\
GR-15310 Athens, Greece.}
\end{center}

\vskip 1 cm

\vskip .5 cm 

\begin{abstract}
We study solutions of the \KZ equation for discrete representations of $SU(2)_k$ at rational level $k+2=\f{p}{q}$ using a regular basis in which the braid matrices are well defined for all spins. We show that at spin $J=(j+1)p-1$ for $2j \in \mathbf{N}$ there are always a subset of $2j+1$ solutions closed under the action of the braid matrices. For $j \in \mathbf{N}$ these fields have integer conformal dimension and all the $2j+1$ solutions are monodromy free. The action of the braid matrices on these can be consistently accounted for by the existence of a multiplet of chiral fields with extra $SU(2)$ quantum numbers ($m=-j,\cdots,j$). In the quantum group $SU_q(2)$, with $q=e^{\f{-i \pi}{k+2}}$, there is an analogous structure and the related representations are trivial with respect to the standard generators but transform in a spin $j$ representation of $SU(2)$ under the extended center.
\end{abstract}

\end{center}

\end{titlepage}

\section{Introduction}
The study of conformal invariance in two dimensions has been a fascinating and productive area of research for the last twenty years \cite{Belavin:1984vu}. The WZNW model is of great importance in CFT. Correlation functions obey differential, Knizhnik-Zamolodchikov (KZ), equations \cite{Knizhnik:1984nr} coming from null states in the theory. For $SU(2)_k$ WZNW model with $k \in \mathbf{N}$ there is a consistent truncation, known as the integrable sector. This involves primary fields with discrete spin $2j \in Z$ in the range $0 \le j \le k/2$ \cite{Gepner:1986wi,Zamolodchikov:1986bd}. The solutions to the KZ equations and correlation functions for these fields were studied in \cite{Zamolodchikov:1986bd,Christe:1987cy}. Using these fields one may construct a closed set of fusion rules and a local modular invariant CFT.

For the case of fractional level $SU(2)_k$ there is a set of fields, known as the admissible representations, which are closed under modular transformations. The correlation functions of such fields have been studied in \cite{Furlan:1991by,Furlan:1993mm,Andreev:1995bj,Furlan:1997vu}. Recently the simplest examples have been carefully re-examined \cite{Gaberdiel:2001ny,Lesage:2002ch} and it was found that under fusion the set of admissible representations is not closed and other (non-admissible) fields are produced. The full consistency of these models is therefore  an open question.

Here, rather than the admissible representations, we shall discuss the discrete ones. It is certainly not the case that a consistent model can be constructed using only these fields. Our primary motivation here is to understand better certain discrete representations that are are monodromy free (i.e. mutually local) and can therefore be considered as extensions of the chiral algebra \cite{Rehren:1995jc,Todorov:2000fm}. We hope that this structure will be useful in understanding models beyond the integrable sector such as the logarithmic conformal field theories (LCFTs) in which indecomposable representations are created in the fusion of irreducible ones \cite{Gurarie:1993xq}. The appearance of logarithms was first observed in the context of WZNW models based on supergroups \cite{Rozansky:1993td,Rozansky:1992rx}. A great deal of work has been done on analysing LCFTs and their applications in many different contexts and references can be found in \cite{Kogan:1997fd,Tabar:2001et,Flohr:2001zs,Gaberdiel:2001tr,Kawai:2002fu,Moghimi-Araghi:2002gk}.

The $c_{p,1}$ models, and in particular the case of $c_{2,1}=-2$, have received considerable attention in the literature due to the discovery that one may extend the Virasoro algebra by triplets of chiral $h_{3,1}=2p-1$ fields \cite{Kausch:1991vg}. In the case of $c_{p,1}$ the characters of the resulting $W$-algebra were shown to provide a finite basis closed under modular transformations \cite{Flohr:1996ea,Flohr:1997vc}. In the case of $c=-2$ it was found that the decoupling of the singular vectors of this algebra gives rise to a local rational LCFT, i.e. one having only a finite number of irreducible and indecomposable representations closed under fusion \cite{Kausch:1995py,Gaberdiel:1996np,Gaberdiel:1998ps,Kausch:2000fu}.

One intriguing aspect is that the extending algebras in all the $c_{p,1}$ cases involved multiplets of chiral operators with extra $SU(2)$ quantum numbers. We have studied many examples of the appearance of such extra multiplet structure in the chiral fields of the $c_{p,q}$ models and also $SU(2)_k$ models (in particular $SU(2)_0$) \cite{Kogan:2001nj,Nichols:2001cv,Nichols:2002dk,Nichols:2002qv}. The first indications of a general pattern were found by studying the appearance of rational correlation functions and their properties under crossing - in effect calculating the braid matrices. However in the case of the $c_{p,q}$ models we were able to give a complete free field representation for these fields making explicit their extra $SU(2)$ quantum numbers \cite{Nichols:2003dj}.

Here we shall take an alternative, and more direct, approach to compute the braid matrices which encode the structure and possible extra quantum numbers of these fields. We shall make use of the regular basis of solutions provided in \cite{Stanev:1992dr,Hadjiivanov:2000gx,Hadjiivanov:2001kr}. We shall first show that for $SU(2)_k$ at rational level $k+2=p/q$ at $J=(j+1)p-1$ the KZ equation always has a subset of $2j+1$ solutions closed under the action of the braid matrices (it may of course also have others). The action of the braid matrices on this set is closed but non-trivial and is exactly as one would expect if the fields also have an \emph{extra} $SU(2)$ quantum number. The integer $j$ fields, whose solutions are monodromy free, always have integer conformal dimension and are therefore good candidates for extensions of the chiral algebra. The way in which the properties of the braid matrices emerge is intimately linked to the quantum group $SU_q(2)$ with $q=e^{\f{-i \pi}{k+2}}$. We show that the corresponding representations in the quantum group are trivial with respect to the standard $SU_q(2)$ generators but transform as a spin $j$ representation of $SU(2)$ under the extended center.
\subsection{\KZ equation}
We shall consider the $SU(2)_k$ affine Kac-Moody algebra:
\bea \label{eqn:SU2KM}
J^3(z) J^{\pm}(w) &\sim& \pm \f{J^{\pm}(w)}{z-w} \nonumber \\
J^+(z)J^-(w) &\sim& \f{k}{(z-w)^2}+\f{2 J^3(w)}{z-w} \\
J^3(z)J^3(w) &\sim& \f{k}{2(z-w)^2} \nonumber
\eea
Affine Kac-Moody primary fields have the following simple OPE:
\bea
J^a(z) \phi_J(w) \sim \f{t^a_{J} \phi_J(z)}{z-w}
\eea
where $t^a_J$ is a spin $J$ matrix representation of $SU(2)$.

One can introduce the following representation for the $SU(2)$ generators \cite{Zamolodchikov:1986bd}:
\bea \label{eqn:repn}
t^+=x^2\frac{\p}{\p x}-2Jx, ~~~
t^-=-\frac{\p}{\p x}, ~~~
t^3=x\frac{\p}{\p x}-J \eea
There is also a similar algebra in terms of $\bar{x}$ for the anti-holomorphic part. It is easily verified that these obey the global $SU(2)$ algebra:
\bea
[t^+,t^-]=2t^3 \quad \quad [t^3,t^{\pm}]=\pm t^{\pm}
\eea
Using these auxiliary variables we can write a discrete representation with \mbox{$M=-J,\cdots, J$} as a single field:
\bea \label{eqn:auxilvar}
\phi_J(x,z)=\sum_{M=-J}^{J} x^{M+J}\phi_{J,M}(z)
\eea
Therefore affine Kac-Moody primary fields $\phi_j(x,z)$ satisfy:
\bea \label{eqn:JphiOPEs}
J^-(z) \phi_J(x,w) &\sim&  -\f{ -\frac{\p}{\p x} \phi_J(x,w) }{z-w} \nonumber\\
J^3(z) \phi_J(x,w) &\sim&  -\f{\left( x\frac{\p}{\p x}-J \right)\phi_J(x,w) }{z-w} \\
J^+(z) \phi_J(x,w) &\sim&  -\f{\left(x^2\frac{\p}{\p x}-2Jx \right)\phi_J(x,w) }{z-w} \nonumber
\eea
The stress tensor is given by the Sugawara construction:
\bea \label{eqn:sugawara}
T&=&\f{1}{2(k+2)}\left( J^+J^- + J^-J^+ + 2 J^3J^3 \right) 
\eea
and the central charge is:
\bea
c=\f{3k}{k+2}
\eea
The fields $\phi_J(x,z)$ are also primary with respect to the Virasoro algebra with conformal weight:
\be
h=\frac{J(J+1)}{k+2}
\ee
The two and three point functions are determined, up to overall scaling, using the global $SU(2)$ and conformal transformations. For the case of the four point correlation functions of $SU(2)$ primaries the form is determined by global conformal and $SU(2)$ transformations only up to a function of the cross ratios. Here we shall only consider correlators with all fields of the same spin:
\bea \label{eqn:correl} 
\langle \phi_{J}(x_1,z_1) \phi_{J}(x_2,z_2) \phi_{J}(x_3,z_3) \phi_{J}(x_4,z_4) \rangle
&=&z_{42}^{-2h} z_{31}^{-2h}x_{42}^{2J} x_{31}^{2J}~F(x,z) 
\eea
where the invariant cross ratios are given by:
\be \label{eqn:ratios}
x=\frac{x_{12}x_{34}}{x_{13}x_{24}} ~~~ z=\frac{z_{12}z_{34}}{z_{13}z_{24}} 
\ee
Correlation functions of the WZNW model satisfy a set of partial differential equations known as  Knizhnik-Zamolodchikov (KZ) equation due to constraints from the null states following from (\ref{eqn:sugawara}). These are:
\be
\left[(k+2) \frac{\p}{\p z_i}+\sum_{j\neq i}\frac{\eta_{ab} J^a_i \otimes J^b_j}{z_i-z_j} \right] \left<\phi_{j_1}(z_1) \cdots \phi_{j_n}(z_n) \right> =0 
\ee
where $k$ is the level of the $SU(2)$ WZNW model. 

For two and three point functions this gives no new information. However for the four point function (\ref{eqn:correl}) it becomes a partial differential equation for $F(x,z)$. For  a compact Lie group this equation can be solved \cite {Knizhnik:1984nr} as it reduces to a set of ordinary differential equations.

If we now use our representation (\ref{eqn:repn}) we find the KZ equation for four point functions is:
\be \label{eqn:KZ}
(k+2) \frac{\p}{\p z} F(x,z)=\left[ \frac{\CP}{z}+\frac{\CQ}{z-1} \right] F(x,z)
\ee
Explicitly these are:
\bea
\CP \!\!\!\!&=&\!\!-x^2(1-x)\frac{\p^2}{\p x^2}+((-4J+1)x^2+4Jx)\frac{\p}{\p x} +4J^2x-2J^2 \\
\CQ \!\!\!\!&=&\!\!-(1-x)^2x\frac{\p^2}{\p x^2} - ((-4J+1)(1-x)^2+4J(1-x))\frac{\p}{\p x} +4J^2(1-x)-2J^2 \nonumber 
\eea
\subsection{Braid matrices}
The effect of exchanging operators in a correlator is encoded within the braid matrices. It is sufficient in considering a correlator of four fields to discuss the effect of the elementary exchanges $1 \leftrightarrow 2$ and $2 \leftrightarrow 3$ which we shall denote by $B_1$ and $B_2$ respectively. In terms of the ratios $x,z$, defined in (\ref{eqn:ratios}), the action is:
\bea
B_1: (x,z) &\rightarrow& \left( \f{x}{x-1},\f{z}{z-1} \right) \nonumber \\
B_2: (x,z) &\rightarrow& \left( \f{1}{x},\f{1}{z} \right)
\eea
These two matrices obey the braid relation:
\bea \label{eqn:braid}
B_1 B_2 B_1=B_2 B_1 B_2
\eea
The action of taking one point completely around another, called monodromy, is given by the square of the braid matrices. Here we shall be interested in possible extensions of the chiral algebra. The correlators of such fields must be rational functions and therefore have trivial monodromy leading to the condition that $B_1^2=B_2^2=1$.

\section{Example: Multiplet structure in $SU(2)_0$}
As an example of the general phenomena that we are discussing in this paper let us review the example of multiplet structure in the chiral fields of $SU(2)_0$ \cite{Nichols:2001cv}. A necessary condition for the existence of a chiral algebra is the
appearance of a rational function, that is a function with only a finite number of poles, within the conformal blocks. The fact that a rational function can be reconstructed from its pole structure and behaviour at infinity is in exact coincidence with the fact that chiral algebras are determined by the singular terms in the OPE. We therefore need only to search for rational solutions to the KZ equation (\ref{eqn:KZ}) for $SU(2)_0$.

For $J=1$ we found one rational solution:
\bea \label{eqn:Jone}
F_{1111}(x,z)=-\f{1}{2(z-1)}+\f{x}{z}+\f{x^2}{2z(z-1)}
\eea
Now the effect of exchange of operators on this correlator is given by:
\bea \label{eqn:braidpropJ1}
F_{1111}(1-x,1-z)&=&F_{1111}(x,z) \\
z^{-2h}x^{2j}F_{1111}\left(\f{1}{x},\f{1}{z}\right)&=&F_{1111}(x,z) \nonumber
\eea
Therefore on this solution the braid matrices are trivial:
\bea \label{eqn:Jonebraid}
B_1=B_2=F=1
\eea
For the four point correlator with $J=2$ we find two rational solutions:
\bea \label{eqn:Jtwo}
F_{2222}^{(1)}(x,z)&=&\f{1}{z^5}\left( (-z^3-2z^4)+(-12z^2-16z^3+4z^4)x+(-18z+18z^3)x^2 \right. \nonumber  \\
&&\quad \left.+(-4+16z+12z^2)x^3+(2+z)x^4 \right) \\
F_{2222}^{(2)}(x,z)&=&\f{1}{(z-1)^5} \left( (10z-15z^2+9z^3-2z^4) \right. \nonumber \\
&&\quad +(60-140z+108z^2-36z^3+4z^4)x \nonumber\\
&&\quad \left.+(-90+162z-90z^2+18z^3)x^2+(36-44z+12z^2)x^3+(-3+z)x^4 \right) \nonumber
\eea
Now we have:
\bea \label{eqn:braidpropJ2}
F_{2222}^{(1)}(1-x,1-z)&=&F_{2222}^{(2)}(x,z)\nonumber \\
F_{2222}^{(2)}(1-x,1-z)&=&F_{2222}^{(1)}(x,z)\\
x^4z^{-6}F_{2222}^{(1)}\left(\f{1}{x},\f{1}{z} \right)&=&-F_{2222}^{(1)}(x,z) \nonumber\\
x^4z^{-6}F_{2222}^{(2)}\left(\f{1}{x},\f{1}{z} \right)&=&-F_{2222}^{(1)}(x,z)+F_{2222}^{(2)}(x,z) \nonumber
\eea
Therefore on the basis $\left( F_{2222}^{(1)},F_{2222}^{(2)} \right)$ the braid matrices are given by:
\bea \label{eqn:Jtwobraid}
B_1=\left(
\ba{cc}
1 & 0 \\
-1 & -1 
\ea
\right) 
\quad 
B_2=\left(
\ba{cc}
-1 & -1 \\
0 & 1 
\ea
\right) 
\quad 
F=\left(
\ba{cc}
0 & 1 \\
1 & 0 
\ea
\right)
\eea
As expected we find $B_1^2=B_2^2=1$ due to the fact that the solutions are rational and therefore monodromy free. 

Upon closer inspection we also find that it is impossible to build a non-trivial chiral correlator transforming in a trivial (i.e one dimensional) representation of the braid group. This is in sharp contrast to the rational solutions occurring within the integrable sector of $k \in \mathbf{N}$ which all transform in trivial representations of the braid group \cite{MST,ST}. Therefore one concludes that it is not consistent to regard these as identical $J=2$ chiral operators in a theory. There are two choices: either there are no chiral operators at $J=2$; or such operators must have some extra quantum numbers and hence the correlators transform in higher dimensional representations of the braid group. By examining the form of the braid matrices (\ref{eqn:Jtwobraid}), or by use of the free field representation \cite{Nichols:2001cv}, one finds that the $J=2$ fields are a fermionic doublet and the correlators are realised in the following way:
\bea
\la \Psi^+ (0,0) \Psi^-(x,z) \Psi^-(1,1) \Psi^+(\infty,\infty) \ra = F_{2222}^{(1)}(x,z) \\
\la \Psi^+ (0,0) \Psi^+(x,z) \Psi^-(1,1) \Psi^-(\infty,\infty) \ra = F_{2222}^{(2)}(x,z) \nonumber
\eea
The deduction of the extra quantum numbers relies only on the action of the braid matrices (\ref{eqn:Jtwobraid}) on the rational correlators and not detailed form of the solutions (\ref{eqn:Jtwo}).
\section{Multiplet structure in the Braid matrices}
\subsection{Regular basis}
In \cite{Stanev:1992dr} a basis of integral solutions to the $SU(2)_k$ KZ equation (\ref{eqn:KZ}) was given. This is called a regular basis as, in contrast to other bases, the relative normalisation of elements remains finite for \emph{all} spins and not just those of the integrable sector. Here we shall only consider the case of four operators with equal spin $J$. The solutions which we denote $w_{\lambda}$ with $\lambda=0,\cdots,2J$, given explicitly in \cite{Stanev:1992dr}, are for the full correlator (\ref{eqn:correl}) including the pre-factor. There is also an obvious $J$ and $k$ dependence of $w_{\lambda}$ which we suppress in the following.

The action of $B_1$ on the regular basis is given by:
\bea
\left(B_1 w \right)_{\mu}&=& \sum_{\lambda=0}^{2J} \left(B_1 \right)^{\lambda}_{\mu} w_{\lambda} \nonumber 
\eea
The properties of the regular basis guarantee that this matrix is well defined and it was found in \cite{Stanev:1992dr} to be given by:
\bea \label{eqn:generalB1}
(B_1)^{\lambda}_{\mu} = q^{2J(k+1-J)} (-1)^{\lambda}  q^{-\lambda (\mu+1)} \qbin{\lambda}{\mu} \quad \quad \lambda,\mu=0,1, \cdots, 2J
\eea
with $q=e^{\f{-\pi i}{k+2}}$. The $q-$binomial coefficient is defined as:
\bea
\qbin{\lambda}{\mu}= \f{[\lambda]!}{[\lambda-\mu]![\mu]!}
\eea
with $[a]!=[a][a-1]!$ with $[0]!=1$. The q-number $[x]$ is defined as:
\bea \label{eqn:qnumber}
[x]=\f{q^x-q^{-x}}{q-q^{-1}}
\eea
The braid matrix $B_2$ is given by:
\bea
B_2=F B_1 F
\eea
where $F$ takes the simple form:
\bea
F^{\lambda}_{\mu}=\delta^{2J+\lambda}_{\mu}= \delta^{2p-2+\lambda}_{\mu}
\eea
with $F^2=1$. The operator $F$ corresponds to the exchange of $1 \leftrightarrow 3$ i.e $(x,z) \leftrightarrow (1-x,1-z)$.
\subsection{Multiplet structure}
We shall now show that there is a particular structure present within certain braid matrices of $SU(2)_k$ at rational values of $k+2=p/q$ with $p,q \in \mathbf{N}$ and $gcd(p,q)=1$.

We first note some simple properties: Firstly as $q=e^{\f{-\pi i}{k+2}}=e^{\f{-\pi i q}{p}}$ we have $q^p=(-1)^q$. For $[x] \notin m \mathbf{Z}$ we have:
\bea \label{eqn:rationotmp}
\f{[x+np]}{[x]}=(-1)^{nq} \quad \quad n \in \mathbf{Z}
\eea
However in the cases $x=m \mathbf{Z}$ this is \emph{not} valid as both the numerator and denominator vanish. In these cases one must take the limit carefully using L'Hopital's rule:
\bea \label{eqn:ratiomp}
\f{[n p]}{[m p]}&=&\lim_{y \rightarrow q} \f{y^{np}-y^{-np}}{y^{mp}-y^{-mp}} \nonumber\\
&=& \lim_{y \rightarrow q} \f{n p}{m p} \f{y^{np-1}-y^{-np-1}}{y^{mp-1}-y^{-mp-1}}\nonumber\\
&=& (-1)^{q(n-m)} \f{n}{m}
\eea
This contains an extra factor of $\f{n}{m}$ compared to the naive application of (\ref{eqn:rationotmp}).

Now let us use these formulae to prove an extremely useful relation\footnote{This result is actually a special case of a more general identity:
\bea \label{eqn:qbin2bingeneral}
\qbin{r_0+p r_1}{s_0+p s_1}=(-1)^{q(r_1-s_1)(s_0+ p s_1)} \qbin{r_0}{s_0} \bin{r_1}{s_1} \quad \quad 0 \le r_0,s_0 < p
\eea}:
\bea \label{eqn:qbin2bin} 
\qbin{(a+1)p-1}{(b+1)p-1} &=& \f{[(a+1)p-1]!}{[(b+1)p-1]![(a-b)p]!} \nonumber \\
&=& \f{[(a+1)p-1] \cdots [(a-b)p+1]}{[(b+1)p-1] \cdots [1]} \nonumber \\
&=& (-1)^{q(a-b)((b+1)p-1-b)} \f{[ap] \cdots [(a-b+1)p]}{[(b)p] \cdots [p]} \nonumber \\
&=& (-1)^{q(a-b)((b+1)p-1)} \f{a}{b} \f{a-1}{b-1} \cdots \f{a-b}{1} \nonumber \\
&=& (-1)^{q(a-b)((b+1)p-1)} \bin{a}{b}
\eea
Motivated by numerical investigations of the appearance of rational solutions in many simple examples \cite{Nichols:2002dk,Nichols:2002qv} we shall consider spins $J=(j+1)p-1$ with $j=\half \mathbf{N}$. We claim that there exists a $2j+1$ dimensional subset of the $w_{\mu}$ vectors, closed under the action of both $B_1$ and $B_2$ and monodromy free\footnote{In the case of $j$ half integer and $p,q$ odd monodromy diagonal with an overall phase}, given by $\mu=(j+m+1)p-1$ with $m=-j,\cdots,j$.

It is instructive to consider explicitly the first few cases in this series.
\subsubsection{Singlet ($j=0$): $J=p-1$}
The $j=0$ case $J=p-1$ (for $q=1$ this case has already been considered in \cite{Hadjiivanov:2001kr} where an explicit solution was also given). Now $\lambda=0,\cdots 2p-2$:
\bea
\qbin{\lambda}{p-1} &=& \f{[\lambda]!}{[\lambda-p+1]! [p-1]!} = \f{[\lambda] \cdots [\lambda-p+2]}{[p-1] \cdots [1]} \nonumber\\
&=& \left\{ \ba{ll} 0 & 0 \le \lambda<p-1 \\
1 & \lambda=p-1\\
\f{[\lambda] \cdots [p] \cdots [\lambda-p+2]}{[p-1] \cdots [1]}=0 & p-1 < \lambda \le 2p-2 \ea \right.
\eea
The last line follows from the fact $[p]=0$ and none of the terms in the denominator vanishes. Therefore the only non-trivial element is when $\lambda=p-1$. Then:
\bea
(B_1 w)_{p-1}=\sum_{\lambda=0}^{2p-2}(B_1)^{\lambda}_{p-1} w_{\lambda} &=& \sum_{\lambda=0}^{2p-2} q^{2J(k+1-J)} (-1)^{(p-1)(q-1)} \qbin{\lambda}{p-1} w_{\lambda} \nonumber\\
&=& q^{2J(k+1-J)}  (-1)^{(p-1)(q-1)} w_{p-1} \nonumber \\
&=& w_{p-1}
\eea
Also note that: $F^{\lambda}_{\mu}=\delta^{2I+\lambda}_{\mu}= \delta^{2p-2+\lambda}_{\mu}$ and so:
\bea
(F w)_{p-1} = \sum_{\lambda=0}^{2p-2} F^{\lambda}_{p-1}w_{\lambda} = w_{p-1} 
\eea
Therefore both the braid matrices $B_1$ and $B_2=F B_1 F$ map $w_{p-1}$ into itself. We define the \emph{reduced} braid matrices to be the restriction of the braid matrices to this basis:
\bea
\tilde{B}_1=\tilde{B}_2=\tilde{F}=1
\eea
We use the tilde to emphasize that it is the reduced braid matrix. This is in exact agreement with the braid matrices of $SU(2)_0$, where $p=2,q=1$, acting on the $J=1$ rational solutions (\ref{eqn:Jonebraid}).
\subsubsection{Doublet ($j=\f{1}{2}$): $J=\f{3}{2}p-1$}
Now we claim an invariant subset of elements is $w_{p-1},w_{2p-1}$. In suppressing the $J$ dependence of $w_{\lambda}$ we must not confuse this $w_{p-1}$ with that of the  previous section as they refer to different spin $J$.

For $\lambda=0,\cdots 3p-2$ we have:
\bea
\qbin{\lambda}{p-1} &=& \f{[\lambda]!}{[\lambda-p+1]! [p-1]!} = \f{[\lambda] \cdots [\lambda-p+2]}{[p-1] \cdots [1]} \nonumber \\
&=& \left\{ \ba{ll} 0 &  0 \le \lambda<p-1 \\
1 & \lambda=p-1\\
0 & p-1 < \lambda < 2p-1 \\
(-1)^{(p-1)q} & \lambda=2p-1 \\
0 & 2p-1 <\lambda \le 3p-2 \ea \right.
\eea
where using (\ref{eqn:qbin2bin}) we have:
\bea
\qbin{2p-1}{p-1}=(-1)^{(p-1)q}\bin{1}{0} = (-1)^{(p-1)q}
\eea
Again for $2p-1 <\lambda \le 3p-2$ we have an uncancelled factor of $[2p]=0$ in the numerator causing it to vanish.

Now:
\bea
\qbin{\lambda}{2p-1}
&=& \left\{ \ba{ll} 
0 & 0 \le \lambda<2p-1 \\
1 & \lambda=2p-1\\
0 & 2p-1 < \lambda \le 3p-2 \ea \right.
\eea
Therefore the matrix $B_1$ maps the set $\{ w_{p-1},w_{2p-1} \}$ into itself. The reduced braid matrix $\tilde{B}_1$ is given by:
\bea \label{eqn:doubletreduced}
\tilde{B}_1=
q^{2J(k+1-J)} \left(
\ba{cc}
1 & 0 \\
(-1)^{p+q} & -1 
\ea
\right)
\eea
Similarly the reduction of $F$, and $B_2$ in this basis is:
\bea
\tilde{F}=\left(
\ba{cc}
0 & 1 \\
1 & 0 
\ea
\right) \hskip 2cm 
\tilde{B}_2=\tilde{F} \tilde{B}_1 \tilde{F}= q^{2J(k+1-J)} \left(
\ba{cc}
-1 & (-1)^{p+q} \\
0 & 1 
\ea
\right)
\eea
These reduced matrices still obey the braid relations (\ref{eqn:braid}) and in addition 
we have $(\tilde{B}_1)^2=(\tilde{B}_2)^2=(-1)^{pq}$. The monodromy matrix is diagonal but with an overall phase factor in the case in which both $p$ and $q$ are odd and in this case the chiral fields are para-fermionic \cite{Fateev:1985mm}. This is in exact correspondence with what was found in specific examples \cite{Nichols:2002dk}.
\subsubsection{Triplet ($j=1$):$J=2p-1$}
This will be the last example that we shall discuss explicitly but it illustrates the subtlety of (\ref{eqn:ratiomp}). We claim that a closed subset of elements in this case is $w_{p-1},w_{2p-1},w_{3p-1}$.

Using similar arguments to before we find:
\bea
\qbin{\lambda}{p-1}=
\left\{ 
\ba{ll} 
0 & 0 \le \lambda<p-1 \\
1 & \lambda=p-1\\
0 & p-1<  \lambda<2p-1\\
(-1)^{(p-1)q} & \lambda=2p-1\\
0 & 2p-1< \lambda<3p-1\\
1 & \lambda=3p-1\\
0 & 3p-1 < \lambda \le 4p-2 
\ea \right.
\eea
Now:
\bea
\qbin{\lambda}{2p-1}=
\left\{ 
\ba{ll} 
0 & 0 \le \lambda<2p-1 \\
1 & \lambda=2p-1\\
0 & 2p-1<  \lambda<3p-1\\
2(-1)^q & \lambda=3p-1\\
0 & 3p-1 < \lambda \le 4p-2 
\ea \right.
\eea
The factor of $2$ is using (\ref{eqn:qbin2bin}):
\bea
\qbin{3p-1}{2p-1}&=& \bin{2}{1} (-1)^q =2(-1)^q
\eea
Finally:
\bea
\qbin{\lambda}{3p-1}=
\left\{ 
\ba{ll} 
0 & 0 \le \lambda<3p-1 \\
1 & \lambda=3p-1\\
0 & 3p-1< \lambda\le 4p-2
\ea \right.
\eea
In this case we find the reduced matrix is:
\bea
\tilde{B}_1&=& q^{2J(k+1-J)} \left( \ba{ccc}
1 & 0 & 0 \\
(-1)^{p+q} & -1 & 0 \\
1 & 2 (-1)^{p+q+1} & 1 
\ea
\right)
\eea
Apart from the pre-factor one can see that the doublet matrix (\ref{eqn:doubletreduced}) is embedded within this. The singlet `matrix', which was just a number $1$, was embedded within the doublet matrix. 

The other reduced matrices are given by:
\bea
\tilde{F}&=& \left( \ba{ccc}
0 & 0 & 1 \\
0 & 1 & 0  \\
1 & 0 & 0 
\ea
\right)
\eea
\bea
\tilde{B}_2&=&\tilde{F} \tilde{B}_1 \tilde{F}= q^{2J(k+1-J)} \left( \ba{ccc}
1 & 2 (-1)^{p+q+1} & 1 \\
0 & -1 & (-1)^{p+q} \\
0 & 0 & 1 
\ea
\right) 
\eea
Again these matrices obey $\tilde{B}_1^2=\tilde{B}_2^2=1$ and so are monodromy free.
\subsubsection{General structure for $J=(j+1)p-1$}
From the previous examples it is simple to see the general structure for $j \in \half N$. We have a closed set of elements $w_{\mu}$ with $\mu=(j+m+1)p-1$ where $m=-j,\cdots, j$. In the entries of $\left(B_1\right)^{\lambda}_{(j+m+1)p-1}$ all are trivial except when $\lambda=(j+M+1)p-1$ where again $M=-j,\cdots, j$ and $M>m$:
\bea
\qbin{(j+M+1)p-1}{(j+m+1)p-1}=(-1)^{q(M-m)((j+m+1)p-1)} \bin{j+M}{j+m}
\eea
The reduced matrix is given by:
\bea
\left( B_1 \right)_m^M=q^{2J(k+1-J)} (-1)^{pq(j+m+1)^2+p(j+M+1)+q(j+M+1)+1} \bin{j+M}{j+m}
\eea
where, for convenience, we drop the tildes and just write this as $B_1$ with the understanding that we are operating on the reduced sub-space.

Now using the congruences modulo $2$:  $a^2 \equiv a$ and $pq\equiv p+q+1$ which hold for $a \in \mathbf{Z}$ and $gcd(p,q)=1$ respectively we can simplify this:
\bea \label{eqn:generalB1simplified}
\left( B_1 \right)_m^M=q^{2J(k+1-J)} (-1)^{j+m+(p+q)(M-m)} \bin{j+M}{j+m}
\eea
Now using the identity\footnote{This is easily proved by multiplying both sides by $t^{j+m}$ and summing over $m=-j,\cdots,j$. The L.H.S. is just the binomial expansion of $\left(1-(1+t)\right)^{j+n}$.}:
\bea
\sum_{M=-j}^{j} (-1)^{j+M} \bin{j+M}{j+m} \bin{j+n}{j+M}= (-1)^{j+n} \delta^{n}_m
\eea
we find:
\bea
\left( B_1^2\right)^n_m=\sum_{M=-j}^j \left( B_1 \right)_m^M \left( B_1 \right)^n_M= q^{4J(k+1-J)} \delta^{n}_m= (-1)^{2jpq}  \delta^{n}_m
\eea
Therefore the monodromy matrix is diagonal and only gets a global phase factor in the case in which $j$ is half integer and both $p$ and $q$ are odd. In this case, as with the $j=\half$, the chiral fields are para-fermionic.

Now $2J-\left((j+m+1)p-1\right)=(j-m+1)p-1$ and so the reduced $F$ matrix is again skew diagonal:
\bea \label{eqn:redF}
F^{m}_{m'}=\delta_{m+m',0}
\eea
%
%
\section{Multiplet structure of the $J=(j+1)p-1$ fields}
We have shown in the previous section that there are sets of monodromy free (or monodromy diagonal) solutions closed under the action of the braid matrices. In terms of the regular basis these are given by $w_{(j+m+1)p-1}$ with $m=-j,\cdots,j$. Here, for simplicity, we shall only consider the case of $q^p=1$ with $p$ odd (i.e. $q$ even) in which case the reduced braid matrix (\ref{eqn:generalB1simplified}) becomes:
\bea \label{eqn:reducedB1pplusqodd}
\left( B_1 \right)_m^M=q^{2J(k+1-J)} (-1)^{j+M} \bin{j+M}{j+m} \quad \quad j+M,j+m=0,\cdots, 2j
\eea
Now we observe that, apart from the pre-factor $q^{2J(k+1-J)}$, this is exactly the same as the $q \rightarrow 1$ limit of the general braid matrix (\ref{eqn:generalB1}) with $\lambda=j+M,\mu=j+m$. As the $q \rightarrow 1$ limit of the quantum group $SU_q(2)$ is just standard $SU(2)$ we conclude that the action of the braid matrix $B_1$ on this basis of solutions is exactly what one would expect if this set of $2j+1$ fields were in a spin $j$ dimensional multiplet of $SU(2)$. The set of solutions $w_{(j+m+1)p-1}$ with $m=-j,\cdots,j$ are simply the conformal blocks of a chiral field at $J=(j+1)p-1$ with $2j+1$ components.

The fact that the fields come with an extra $SU(2)$ index suggests that we introduce \emph{another} auxiliary variable $y$ to combine all the chiral multiplet fields with $m=-j,\cdots,j$ into a single field:
\bea
\Phi(x,z;y)=\sum_{m=-j}^{j} y^{j+m} \Phi_m(x,z)
\eea
The global generators are:
\bea
W^+=y^2\frac{\p}{\p y}-2jy, ~~~ 
W^-=-\frac{\p}{\p y}, ~~~
W^3=y\frac{\p}{\p y}-j 
\eea
Now in an exactly analogous way to the global $SU(2)$ Ward identities for the $x$ variables we can write the two and three point functions with the additional $y$ variable:
\be \label{eqn:2pt}
\la \phi_{J_1,j_1}(x_1,z_1;y_1) \phi_{J_2,j_2}(x_2,z_2;y_2) \ra = A(j_1,J_1) \delta_{j_1 j_2} \delta_{J_1,J_2} x_{12}^{2J_1}y_{12}^{2j_1}z_{12}^{-2h}
\ee
\bea \label{eqn:3pt}
\la \phi_{J_1,j_1}(x_1,z_1;y_1)\phi_{J_2,j_2}(x_2,z_2;y_2)\phi_{J_3,j_3}(x_3,z_3;y_3) \ra &=& C(J_1,J_2,J_3;j_1,j_2,j_3)\nonumber\\
&& x_{12}^{J_1+J_2-J_3} x_{13}^{J_1+J_3-J_2} x_{23}^{J_2+J_3-J_1} \\
&& y_{12}^{j_1+j_2-j_3} y_{13}^{j_1+j_3-j_2} y_{23}^{j_2+j_3-j_1} \nonumber\\
&& z_{12}^{-h_1-h_2+h_3} z_{13}^{-h_1-h_3+h_2} z_{23}^{-h_2-h_3+h_1} \nonumber
\eea
The four point function as before is only determined up to a functional form by global properties (again we consider operators with same spin $J$ and isospin $j$):
\bea \label{eqn:4pt}
\langle \phi_{j,J}(x_1,z_1;y_1) \phi_{j,J}(x_2,z_2;y_2) \phi_{j,J}(x_3,z_3;y_3) \phi_{j,J}(x_4,z_4;y_4) \rangle
&=&z_{42}^{-2h} z_{31}^{-2h}x_{42}^{2J} x_{31}^{2J} 
\\ && \quad y_{42}^{2j} y_{31}^{2j}~F(x,z;y) \nonumber
\eea
where now we have the additional invariant cross ratio:
\bea
y=\frac{y_{12}y_{34}}{y_{13}y_{24}}
\eea
These correlation functions now transform in a one dimensional representation of the braid group. It is clear that $F(x,z;y)$ still obeys the KZ equation as this does not act on the $y$ variables. Therefore $F(x,z;y)$ should be written as a linear combination of the $w_{(j+m+1)p-1}$ with each term multiplied by some polynomial in $y$, of maximum degree $2j$, such that the entire expression (with all auxiliary variables) transforms in a one dimensional representation under the action of the braid group. 

In the simple example of the $J=2$ chiral fields from $SU(2)_0$ this is in fact not sufficient to determine the overall form. There are two possibilities transforming in a one dimensional representation. However we know that the fusion rules of the doublet field takes the form \cite{Nichols:2001cv}:
\bea
\Psi^{\alpha} \otimes \Psi^{\beta} = d^{\alpha \beta} [0] + t^{\alpha \beta}_a [W^a]
\eea
and therefore as $y \rightarrow 0$ the function should go as $O(z^0)$ reflecting the contribution of the triplet field. Only one of the solutions has this behaviour and therefore we conclude that the correct correlator is given by:
\bea
F(x,z;y)= y F_{2222}^{(1)}(x,z) + (1-y) F_{2222}^{(2)}(x,z)
\eea
with:
\bea
F(1-x,1-z;1-y)&=&F(1-x,1-z;1-y) \nonumber \\
z^{-6}x^{4} y F \left( \f{1}{x},\f{1}{z};\f{1}{y} \right)&=&-F(x,z;y)
\eea
It is not clear in general what additional conditions govern the consistent fusion of these operators giving the correct $y$ dependence in $F(x,z;y)$.
\section{Quantum group $SU_q(2)$}
In this section we wish to show that in the quantum group $SU_q(2)$ representations with $J=(j+1)p-1$ possess some quite remarkable properties. A general discussion of representations has been given in \cite{Pasquier:1990kd} however we shall analyse separately the representations at $J=(j+1)p-1$ to emphasize the extra structure. For simplicity\footnote{The other cases follow in a similar way but some extra re-definitions are required to eliminate tedious factors of $(-1)$.} we shall take $q^p=1$ with $p$ odd.

We take the standard $SU_q(2)$ relations (see \cite{Faddeev:1990ih} and references therein):
\bea
q^H S^{\pm} = S^{\pm} q^{H \pm 1} \quad \quad 
\left[ S^+, S^- \right] = \f{q^{2H}-q^{-2H}}{q-q^{-1}}
\eea
$SU_q(2)$ can be given a Hopf algebra structure \cite{Jimbo:1985zk,Drinfeld:1985rx} with co-product:
\bea \label{eqn:coproduct}
\Delta\left(q^{\pm H}\right)=q^{\pm H} \otimes q^{\pm H} \quad \quad \Delta\left(S^{\pm}\right)=q^{H} \otimes S^{\pm} + S^{\pm} \otimes q^{-H}
\eea
Let us consider a highest state $\state{J;J}$ acted on by operators $S^-$. We have the convention:
\bea \label{eqn:highlow}
S^{\pm} \state{J; \pm J}=0 \quad \quad \f{\left(S^{\pm} \right)^p}{[p]!}\state{J; \pm J}=0
\eea
One can then show the relation:
\bea \label{eqn:commutation}
\left[ \left(S^+ \right)^n, \left(S^- \right)^m \right] \state{J;J}= \f{[m]!}{[m-n]!} \left(S^- \right)^{m-n} \prod_{k=1}^n \left(\f{q^{2H-m+k}-q^{-2H+m-k}}{q-q^{-1}}\right) \state{J;J}
\eea
valid for $m \ge n$.

Now let us consider the possibility of states $\state{Y}$ which are stationary under the action of $S^{\pm}$:
\bea  \label{eqn:stationarystates}
S^+ \state{Y}=0 \quad \quad S^- \state{Y}=0
\eea
The state:
\bea \label{eqn:stateX}
\state{X}=\f{(S^-)^{p-1}}{[p-1]!} \state{J;J}
\eea
certainly obeys $S^- \state{X}=(S^-)^p \state{J;J}=0$. Acting by $S^+$ gives:
\bea
S^+ (S^-)^{p-1} \state{J;J}= \left[ S^+ , \f{\left(S^- \right)^{p-1}}{[p-1]!} \right] \state{J;J}= \f{\left(S^- \right)^{p-2}}{[p-2]!} [2J-(p-1)+1]\state{J;J}
\eea
Therefore this will also vanish if $2J-(p-1)+1=np$ for $n \in Z$ in other words $J=\f{n+1}{2}p-1$. The fact that we are discussing discrete representations, with $J \in \half \mathbf{N}$, and that we have not already hit the lowest state $\state{J;-J}$ means $n+1 \in \mathbf{N}$. Defining $j$, where $2j \in \mathbf{N}$, by $n=2j+1$ we have $J=(j+1)p-1$.

In \cite{Pasquier:1990kd} the spectrum was explicitly truncated to the cohomology:
\bea \label{eqn:cohomol}
\f{Ker S^+}{Im \left(S^+ \right)^{p-1}}
\eea
This removes all indecomposable representations, and their associated zero norm states, in order to get a `physical' spectrum. In models related to LCFT this truncation is clearly not appropriate as such zero norm states are a natural feature \cite{Caux:1996nm}.

Although the action of the generators $S^{\pm}$ on $\state{X}$ is trivial one may generate a non-trivial action with the elements:
\bea
s^{\pm}=\f{\left( S^{\pm}\right)^p}{[p]!}
\eea
To see this let us compute:
\bea
s^- \state{X} = \f{\left( S^{-}\right)^p}{[p]!} \f{(S^-)^{p-1}}{[p-1]!}\state{J;J} = \qbin{2p-1}{p-1} \f{(S^-)^{2p-1}}{[2p-1]!} \state{J;J} =\f{(S^-)^{2p-1}}{[2p-1]!}  \state{J;J}
\eea
%
In a similar way to before the resulting state is again annihilated by $S^{\pm}$ but the $s^{\pm}$ have a non-trivial action. By continued action of $s^-$ we generate a series of states of the form:
\bea \label{eqn:genstate}
\state{\Phi_{j;m}}=\f{(S^-)^{(j-m+1)p-1}}{[(j-m+1)p-1]!} \state{J;J}
\eea
where we have introduced the parameter $m=j,j-1,\cdots$. Due to the condition (\ref{eqn:highlow}) the final one is when $m=-j$ and therefore we have $m=-j,\cdots,j$. The action of $s^-$ on a general state (\ref{eqn:genstate}) is given in general by:
\bea
s^- \state{\Phi_{j;m}} &=& \f{\left( S^{-}\right)^p}{[p]!} \f{(S^-)^{(j-m+1)p-1}}{[(j-m+1)p-1]!}\state{J;J} \nonumber\\
&=&  \qbin{(j-m+2)p-1}{(j-m+1)p-1} \f{(S^-)^{(j-m+2)p-1}}{[(j-m+2)p-1]!} \state{J;J} \nonumber \\
&=& \bin{j-m+1}{j-m}\state{\Phi_{j;m-1}} \nonumber\\
&=&(j-m+1) \state{\Phi_{j;m-1}}
\eea
Now using (\ref{eqn:commutation}) we find:
\bea
s^+ \state{\Phi_{j;m}} &=& \f{\left( S^{+}\right)^p}{[p]!} \f{(S^-)^{(j-m+1)p-1}}{[(j-m+1)p-1]!}\state{J;J}   \nonumber\\
&=& \f{(S^-)^{(j-m)p-1}}{[(j-m)p-1]!} \f{1}{[p]!}\prod_{k=1}^p \left[ 2J-((j-m+1)p-1)+k \right] \state{J;J} \nonumber\\
&=& \f{(S^-)^{(j-m)p-1}}{[(j-m)p-1]!} \f{1}{[p]!}\prod_{k=1}^p \left[ (j+m+1)p+k-1\right] \state{J;J} \nonumber\\
&=& \qbin{(j+m+1)p+p-1}{(j+m+1)p-1} \f{(S^-)^{(j-m)p-1}}{[(j-m)p-1]!} \state{J;J} \nonumber\\
&=& (j+m+1) \state{\Phi_{j;m+1}}
\eea
where again due to the condition (\ref{eqn:highlow}) we have:
\bea
s^+ \state{\Phi_{j;j}}=0
\eea
In summary we have:
\bea
S^{\pm} \state{\Phi_{j;m}} &=& 0 \nonumber \\
s^{\pm} \state{\Phi_{j;m}} &=& (j \pm m+1) \state{\Phi_{j;m \pm 1}}
\eea
with:
\bea
s^+ \state{\Phi_{j;j}} =0  \quad \quad s^- \state{\Phi_{j;-j}}= 0
\eea
Therefore the states $\state{\Phi_{j;m}}$ are trivial with respect to the generators $S^{\pm}$ but transform as a spin $j$ representation of $SU(2)$ with respect to the central elements $s^{\pm}$.

It is natural to define the operator:
\bea
s^3= \f{1}{2} \f{1}{[p]!} \prod_{k=1}^p \f{q^{2H-p+1}-q^{-2H+p-1}}{q-q^{-1}}=\f{1}{2} \qbin{2H}{p}
\eea
so that on states in $Ker S^+$:
\bea
\left[ s^+,s^- \right]=2 s^3
\eea
The action on the states is:
\bea
s^3 \state{\Phi_{j;m}} &=& \f{1}{2} \qbin{2mp}{p} \state{\Phi_{j;m}} =m \state{\Phi_{j;m}}
\eea
One may also consider the tensor product of two multiplets $\Phi_{j_1}$ and $\Phi_{j_2}$. %
\bea \label{eqn:tensorprodU}
\Phi_{j_1} \otimes \Phi_{j_2}= \sum_{i} U_i
\eea
Now using the co-product formula (\ref{eqn:coproduct}) we have:
\bea
S^{\pm} (\Phi_{j_1} \otimes \Phi_{j_2})= q^H  \Phi_{j_1} \otimes S^{\pm}\Phi_{j_2} +  S^{\pm}\Phi_{j_1} \otimes q^{-H} \Phi_{j_2}=0
\eea
and therefore all the terms $U_i$ in the sum (\ref{eqn:tensorprodU}) are annihilated by $S^{\pm}$ and so are also operators of the form $\Phi_{j;m}$. Also we have:
\bea
s^{\pm} (\Phi_{j_1} \otimes \Phi_{j_2})&=&\f{\left( S^{\pm}\right)^p}{[p]!}  (\Phi_{j_1} \otimes \Phi_{j_2}) \nonumber\\
&=& \Phi_{j_1} \otimes \f{\left( S^{\pm}\right)^p}{[p]!}\Phi_{j_2}+ \f{\left( S^{\pm}\right)^p}{[p]!}\Phi_{j_1} \otimes \Phi_{j_2}  \nonumber \\
&=& \Phi_{j_1} \otimes s^{\pm} \Phi_{j_2}+ s^{\pm} \Phi_{j_1} \otimes \Phi_{j_2}  
\eea
where in between the first and second lines we have used repeatedly the fact that the $\Phi_j$ are annihilated by $S^{\pm}$ and so the only non-trivial terms are when all the $p$ operators act on the same term. 

The action of this on tensor products of the $\Phi_{j;m}$ operators is easily calculated:
\bea
[s^+,s^-] \left(\Phi_{j_1} \otimes \Phi_{j_2}\right) &=&
 s^+ \left( \Phi_{j_1} \otimes s^-\Phi_{j_2} + s^-\Phi_{j_1} \otimes \Phi_{j_2} \right) \nonumber\\
 && \quad - s^- \left( \Phi_{j_1} \otimes s^+\Phi_{j_2} + s^+\Phi_{j_1} \otimes \Phi_{j_2}\right)\nonumber\\
&=& \Phi_{j_1} \otimes [s^+,s^-] \Phi_{j_2}+ [s^+,s^-]\Phi_{j_1} \otimes \Phi_{j_2} 
\eea
as the cross terms cancel. Therefore the action of $s^3$ on tensor products of $\Phi_{j;m}$ is given by:
\bea
s^3 \left(\Phi_{j_1} \otimes \Phi_{j_2}\right) = s^3\Phi_{j_1} \otimes \Phi_{j_2} + \Phi_{j_1} \otimes s^3 \Phi_{j_2}
\eea
and so the quantum numbers of $s^3$, namely $m$, behave additively in the tensor product just as in normal $SU(2)$. Therefore the tensor products of these fields are simply given by:
\bea
\Phi_{j_1} \otimes \Phi_{j_2}= \sum_{j=|j_1-j_2|}^{j_1+j_2} \Phi_j
\eea
and hence they form a closed subset with standard $SU(2)$ tensor product rules. This is in exact coincidence with the fusion rules of the extended chiral multiplets occurring in the $c_{p,q}$ models based on a free field construction \cite{Nichols:2003dj}. It is also clear that by tensoring the $\Phi_j$ with other representations of the quantum group one can give multiplet structure to them. The fields $\Phi_j$ are fundamental in the sense that they are annihilated by both the $S^{\pm}$ - a property only shared by the identity.

The appearance of a non-trivial action of the central elements is extremely similar to the restricted specialization of quantum groups at the roots of unity \cite{ChariPressleyBook}. In the case of $SU_q(2)$, rather than generators of the form $S^{\pm}$, one considers $\left( S^{\pm} \right)^{(n)}=\f{\left( S^{\pm} \right)^n}{[n]!}$. Using (\ref{eqn:qbin2bingeneral}) one can easily derive the formula:
\bea
\left( S^{\pm} \right)^{(r_0+p r_1)}=\left( S^{\pm} \right)^{(r_0)} \f{\left( \left( S^{\pm} \right)^{(p)} \right)^{r_1}}{r_1!}
\eea
showing that all elements can be multiplicatively generated from $S^{\pm}$ and $s^{\pm}=\left( S^{\pm}\right)^{(p)} $. It would be interesting to explore more deeply the relationship between quantum groups and LCFT.
\section{Coulomb gas description}
The free field representation of $SU(2)_k$ parallels that of the minimal models \cite{Dotsenko:1984nm,Dotsenko:1990ui,Dotsenko:1991zb}. We use a bosonic ghost system $(\beta,\gamma)$ and free boson $\phi$ obeying the standard free field relations:
\bea
\beta(z) \gamma(w) \sim \f{-1}{z-w} \quad  \quad  \phi(z) \phi(w) \sim -\ln(z-w) 
\eea
In the Wakimoto representation \cite{Wakimoto:1986gf} the $SU(2)_k$ currents (\ref{eqn:SU2KM}) are realised as:
\bea \label{eqn:Wakimoto}
J^+&=&\beta \nonumber \\
J^3&=&i \sqrt{\f{k+2}{2}} \p \phi +  \gamma \beta\\ 
J^-&=&-i \sqrt{2(k+2)} \p \phi \gamma - k \p \gamma - \beta\gamma^2 \nonumber
\eea
The Sugawara stress tensor (\ref{eqn:sugawara}) is:
\bea
T&=& -\beta \p \gamma - \half \p \phi \p \phi - \f{i}{\sqrt{2(k+2)}} \p^2 \phi
\eea
The central charge is made up by: $2+\left(1- \f{6}{k+2} \right)=\f{3k}{k+2}$.

In the free field representation one introduces screening operators $Q$:
\bea
Q=\oint J(z) dz= \oint dz \beta e^{-i \sqrt{2} ~\alpha_+ \phi} \quad \quad \quad \alpha_+=\sqrt{\f{1}{k+2}}=\sqrt{\f{q}{p}}
\eea
Primary operators are given by vertex operators of the form:
\bea
V_{J;M}=\left(-\gamma \right)^{J-M} e^{i \sqrt{2} j \alpha_+ \phi}
\eea
In correlators one must insert the screening charges to ensure the overall charge neutrality of the correlator. In the Coulomb gas description of $SU(2)_k$ the quantum group structure is realised with these screening charges acting as lowering operators \cite{Gomez:1990er,Gomez:1991sw}. The quantum group expression for the multiplet fields given in (\ref{eqn:genstate}) suggests the following combination:
\bea \label{eqn:Wffrepn}
W_{j;m}(z)&=& \f{1}{[(j+m+1)p-1]!} \times  \nonumber \\
&&\oint d{z_1} \cdots \oint d{z_{(j+m+1)p-1}}  J(z_1) \cdots J(z_{(j+m+1)p-1}) e^{i \sqrt{2} \left((j+1)p-1\right) \alpha_+ \phi(z)}
\eea
Inside correlation functions this certainly can make sense. It is well known in LCFT that such divergent normalisations must be taken in order get non-trivial finite results. The properties that we proved in the previous section essentially allow $W_{j;m}(z)$ to be inserted into arbitrary correlators without disturbing the other screening integrals. In fact (\ref{eqn:Wffrepn}) is a generalisation of the expressions given in \cite{Kausch:1991vg} for the case $p=1$.

However as an operator it is not quite clear how to make precise sense of (\ref{eqn:Wffrepn}). On examination one can see that (\ref{eqn:Wffrepn}) will be a descendent of the vertex operator $e^{- i \sqrt{2pq}m \phi}$ in exact agreement with what we previously found in the free field representation of similar extending algebras in the $c_{p,q}$ models \cite{Nichols:2003dj}. However in the free field representation of \cite{Nichols:2003dj} we found that one must also involve the logarithmic partner fields of \cite{Flohr:1997wm} in order to get non-vanishing results. It seems that the appearance of logarithmic structure is intimately related to the existence of these extended algebras (and vice-versa). A description of LCFTs using an extended state space has been proposed in \cite{Fjelstad:2002ei} and it would be interesting to investigate the relation between this and the extending algebras.
\section{Conclusion}
We have investigated the extra chiral multiplet fields occurring at $J=(j+1)p-1$ in $SU(2)_k$ at level $k+2=p/q$. We have shown that there are $2j+1$ solutions on which the action of the braid matrices is closed but non-trivial. For integer $j$ these are always monodromy free. The occurrence of such structure gives rise to an extra set of $SU(2)$ quantum numbers in the theory.

In the quantum group $SU_q(2)$ we have shown that the corresponding representations have the property that they are trivial under the action of $S^{\pm}$. If we were to regard the generators $S^{\pm},q^H$ of $SU_q(2)$ as `local' then the representations that we have found are \emph{topological}. The extra $SU(2)$ quantum numbers are simply charges under the extended center of $SU_q(2)$. The tensor product of these representations is closed and governed by standard $SU(2)$ rules. It seems likely that the structure we have given here for $SU_q(2)$ will generalise to any quantum group at a root of unity with the extended center acting as a global group.

Although we have emphasized the applications of our results to LCFT there is one important non-logarithmic example: the non-minimal $c_{1,1}=1$ model. This correponds to a free boson at self-dual radius which possesses additional $SU(2)$ structure, in this case simply $SU(2)$ affine Kac-Moody algebra \cite{Ginsparg:1988eb}. It is pleasing to see how the extended structure present there is just a special case of a much more general phenomena.

There are many issues which deserve further investigation. We have seen hints of a deeper connection between LCFT and quantum groups at the roots of unity. It would be interesting to see how much of the structure of LCFT, for instance operator content and fusion rules, could be deduced from the corresponding quantum group. It would also be interesting to learn how to generalise the Coulomb gas description of minimal models \cite{Felder:1989zp} to the LCFT case. Given that the chiral multiplet algebras are crucial in producing a rational LCFT it would be surprising if they did not play a correspondingly important role in integrable models beyond the truncated sector. Hopefully this will shed some light on the \emph{physical} meaning of these symmetries. 
\section{Acknowledgements}
I am grateful to the EU for funding under the {\it Discrete Random Geometries: From Solid State Physics to Quantum Gravity} network.
%
%

\end{document}